\documentclass{article}

\usepackage{bioarxiv}
\usepackage[utf8]{inputenc} 
\usepackage[T1]{fontenc}    
\usepackage{hyperref}       
\usepackage{url}            
\usepackage{booktabs}       
\usepackage{amsfonts}       
\usepackage{nicefrac}       
\usepackage{microtype}      
\usepackage{graphicx}
\usepackage{wrapfig}
\usepackage{amsmath}
\usepackage{multicol}        
\usepackage[bottom]{footmisc}
\usepackage{newtxtext}   
\usepackage{newtxmath}
\usepackage{makecell}
\usepackage{adjustbox}

\usepackage{subfig}
\usepackage{tabularx}
\usepackage{longtable}
\usepackage{multirow}
\usepackage{tabularx}
\usepackage{lscape}
\usepackage{setspace}

\title{Twelve quick tips for designing AI-driven HPC workflows}

\author{
  Jamie J. Alnasir\\
  Department of Computer Science\\
  Royal Holloway University of London\\
  Egham, England, UK \\
  \texttt{jamie.alnasir@rhul.ac.uk} \\
}

\begin{document}

\maketitle



\section*{Introduction}
High-performance computing (HPC) clusters remain the backbone of large-scale scientific computation, allowing researchers to run resource-intensive workloads across distributed systems. Historically, these applications relied on deterministic, linear pipelines that processed raw input data through an unchanging series of sequential or parallel execution steps.

The rapid adoption of artificial intelligence (AI) and machine learning (ML) models across the life sciences has disrupted this traditional model. Unlike static pipelines, modern AI-driven configurations are inherently iterative, data-dependent, and probabilistic. They introduce structural cycles of model training, batch inference, validation, and automated refinement\mbox{~\cite{ejarque2022hpcai}}.

\begin{figure}
\centering
\includegraphics[width=\textwidth,height=0.70\textheight,keepaspectratio]{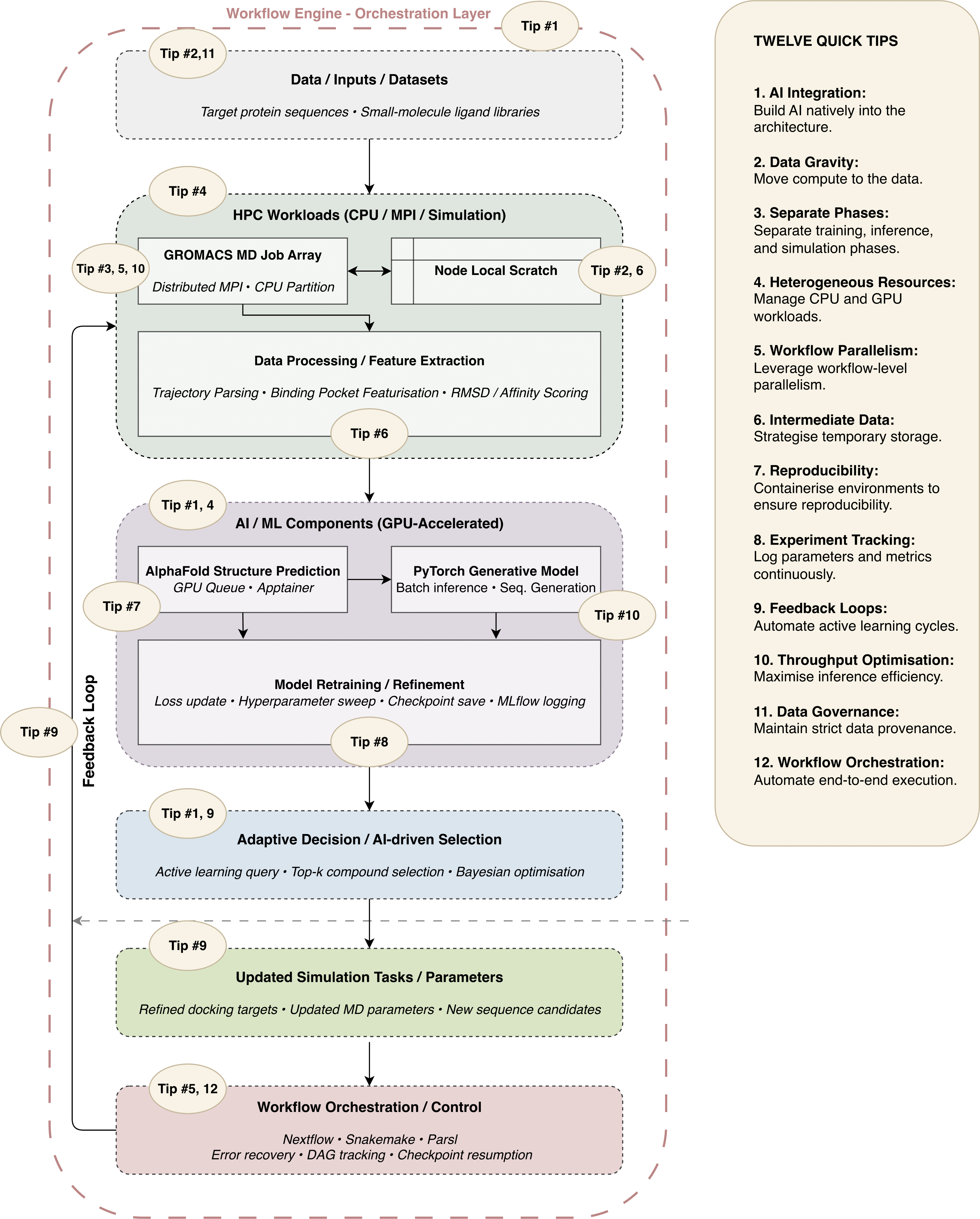}
\caption{
Example architecture for an AI-enabled HPC workflow with an adaptive
feedback loop in computational biology. The figure illustrates an active-learning workflow for structure-guided protein or small-molecule design. Input datasets, including target protein sequences and ligand libraries, are processed through CPU/MPI simulation tasks, GPU-accelerated machine-learning components, and workflow-engine orchestration. Intermediate outputs such as docking scores, molecular-dynamics trajectories, embeddings, and model checkpoints are written to appropriate storage tiers and reused in subsequent stages. The adaptive feedback loop allows simulation and model outputs to guide the next round of candidate selection, while containers, provenance logging, scheduler integration, and data-governance controls support reproducibility and operational robustness. Tip numbers indicate where each recommendation applies within the workflow.
}
\label{fig:ai_hpc_workflow}
\end{figure}

This operational shift changes the way AI-enabled workloads use cluster resources. Traditional HPC applications often have predictable control flow, uniform node usage and relatively stable resource requirements. By contrast, AI-enabled workflows may adapt at runtime, with intermediate results influencing task selection, resource usage, and data movement. This means their runtime structures, resource priorities, and data routing may need to adapt dynamically based on intermediate statistical inferences. Managing this transition requires an intentional shift in how workloads are designed, orchestrated, and scaled. This conceptual shift is illustrated in Fig~\ref{fig:ai_hpc_workflow}.

This guide provides twelve quick tips for research software engineers (RSEs), workflow developers, and bioinformaticians who design, maintain, or scale AI-driven HPC workflows in the life sciences. The emphasis is on practical workflow design principles that support robust and reproducible execution across traditional HPC and AI-enabled computational pipelines. It builds on earlier practical guidance for using HPC systems effectively, while shifting the focus from general HPC usage to the design challenges introduced by AI-driven and adaptive workflows\mbox{~\cite{alnasir2021hpc}}.

Although the principles are broadly applicable across scientific computing, they are especially relevant to computational biology workflows involving large-scale simulation, imaging, omics analysis, model training, and iterative data-driven discovery. They are equally applicable to emerging workflows involving foundation models, large-scale inference services, and adaptive AI systems that increasingly interact with traditional computational biology pipelines.

These emerging systems also impose distinct systems-level constraints: large-scale model training may require distributed GPU execution using combinations of data, tensor, and pipeline parallelism; inference serving may be limited by batching and GPU-memory management; and checkpointing large model states can create substantial I/O pressure if handled naively~\cite{narayanan2021megatron,kwon2023pagedattention,maurya2024datastates}.

\section*{Tip 1: Integrate AI components explicitly into the workflow}

Adding machine learning to an HPC pipeline as a standalone scoring step is quick, but it often produces fragile workflows because model behaviour depends on training data, architecture, hyperparameters, random seed initialisation, software environment, and model version.

Treat the model as a named workflow component with explicit inputs, outputs, versions, and validation points. The workflow should define how training or update data are produced, how probabilistic predictions are consumed by downstream tasks, and how validation results are returned to subsequent runs.

This makes provenance and control flow auditable rather than leaving them in ad hoc wrapper scripts. In a virtual-screening workflow for viral inhibitors, inference scores should not simply be passed to molecular dynamics (MD). The workflow should record the model and feature set used, route selected compounds or conformations to docking or MD validation, and use the resulting scores to expand, prune, or retrain the active-learning loop.

\section*{Tip 2: Design for ``data gravity'', not just compute locality}

AI-enabled workflows are often constrained by data movement and storage throughput as much as compute. Repeatedly reading large training datasets from shared parallel filesystems can saturate interconnects, metadata servers, and storage back ends, reducing throughput for both the workflow and other users.

Treat data layout as part of workflow design. Chunked formats can reduce metadata pressure and support parallel reads more effectively than large collections of small files. In bioimaging, OME-Zarr supports scalable access to multidimensional image data~\cite{moore2023omezarr}; for broader HPC workflows, ADIOS2 provides high-performance data management for intermediate and streaming data~\cite{godoy2020adios2}.

Design for ``data gravity'' by bringing computation to the data, rather than repeatedly moving data to computation. This means storing primary datasets on shared or high-performance storage accessible to compute nodes, staging frequently accessed data onto node-local scratch where appropriate, and minimising redundant transfers between workflow steps.

Specify the direction and frequency of data movement between AI and HPC components, since repeated bidirectional exchange can dominate tightly coupled workflows. For example, a drug-response model trained from whole-genome sequencing data should avoid streaming multi-terabyte BAM or FASTQ-derived inputs from the shared filesystem at every epoch; converted or chunked training arrays can be staged once to node-local scratch and read locally during training.

\section*{Tip 3: Separate training, inference, and simulation phases}

AI-driven workflows often combine phases with different computational profiles. Training is stateful and may require sustained GPU use, high memory, and long wall-clock allocations. Inference is usually lighter and more easily parallelised across independent inputs. Simulation or validation stages may instead require CPU nodes, MPI execution, or long-running batch jobs.

Treating these phases as equivalent can produce inefficient resource requests, idle accelerators, and avoidable queue pressure. Training jobs should be submitted to GPU-enabled queues with appropriate resource allocations, inference tasks can be distributed across available nodes using job arrays or similar mechanisms, and simulation steps should be scheduled independently using resources suited to their requirements.

This distinction is increasingly important for foundation-model-style workloads, where inference may be constrained by batching, request scheduling, and key-value-cache memory management rather than by training bottlenecks~\cite{kwon2023pagedattention}. In protein-binder design, for example, molecular dynamics trajectories can run on CPU/MPI resources while generative-model training or binder inference is invoked on GPU nodes only when suitable batches are ready. Separating these phases makes resource use explicit and prepares the workflow for the heterogeneous resource mapping discussed in Tip 4.

\section*{Tip 4: Use heterogeneous resources intentionally}

AI-HPC workflows rarely need the same hardware throughout. Preprocessing, filtering, and postprocessing are often CPU-bound; model training and accelerator-suitable inference benefit from GPUs; and unusually large models or intermediate datasets may require high-memory nodes. Requesting GPU resources for every stage wastes specialised allocations and can increase queue times.

Specify resources per task rather than per workflow. Workflow engines and scheduler directives should state CPU, memory, accelerator, wall-time, and storage requirements for each stage, including tasks placed on remote or distributed systems. This helps schedulers place work appropriately and avoids holding scarce GPUs idle while CPU-bound steps run.

A cryo-EM pipeline illustrates the issue: motion correction, filtering, and particle picking may run efficiently on CPU nodes, while neural-network-based 3D reconstruction should move to GPU nodes only after suitable particle data have been prepared. Mapping stages to resource types improves utilisation and reduces contention on shared systems.

\section*{Tip 5: Leverage workflow-level parallelism}

Many AI-HPC workflows contain large numbers of independent tasks, including sample-level preprocessing, batch inference, docking jobs, and hyperparameter evaluations. These should be expressed as workflow-level parallelism rather than serial shell loops. Suitable mechanisms include scheduler job arrays, workflow-engine scatter/gather patterns, and task-parallel frameworks that expose independent units of work while retaining dependencies and provenance.

Task granularity should be chosen deliberately. Very small tasks can overwhelm schedulers or workflow runtimes, whereas very large tasks reduce scheduling flexibility and make failures more expensive. Batch inputs so that each task has enough work to amortise launch overhead while allowing failed or slow tasks to be retried locally.

In virtual screening, for example, a ligand library can be divided into scheduler-managed batches. Each task can run docking across available CPU cores and pass selected poses to GPU-based scoring where appropriate. This is more scalable than a monolithic loop and allows throughput to increase as cluster slots become available.

\section*{Tip 6: Manage intermediate data and state}

AI workflows generate substantial intermediate state, including preprocessed inputs, temporary feature matrices, checkpoints, logs, and model artefacts. If unmanaged, these files can exhaust quotas, overload shared filesystems, or prevent recovery after an interrupted job.

Manage state deliberately. Use node-local scratch for temporary files, retain only outputs required for audit or reuse, checkpoint long-running training jobs, and clean up transient files automatically. Where steps are submitted directly to a batch scheduler, job dependencies can express continuation or recovery logic, such as launching validation or restart jobs only after a checkpoint-producing stage completes.

Checkpointing should balance resilience against I/O overhead. Large neural-model checkpoints may include sharded parameters, optimiser states, and runtime metadata; writing them synchronously to a shared parallel filesystem can become a bottleneck, motivating asynchronous or multi-level checkpointing and high-performance data-management approaches~\cite{maurya2024datastates,godoy2020adios2}.

A single-cell RNA-seq workflow illustrates the trade-off. Temporary filtering arrays and intermediate expression matrices can be written to node-local scratch, while final embeddings, model weights, and provenance records are retained on persistent storage. This reduces shared-filesystem pressure while preserving the state needed to audit, resume, or reproduce the workflow.

\section*{Tip 7: Containerise environments to support reproducibility}

AI-HPC workflows often depend on tightly coupled software stacks: Python packages, CUDA or ROCm libraries, compiler versions, MPI implementations, model weights, and domain tools. Rebuilding these environments manually on shared systems is fragile, especially when modules or drivers change.

Use container images to define the execution environment explicitly. On HPC systems, Apptainer and SingularityCE are commonly used because they integrate with schedulers and do not normally require elevated privileges at runtime~\cite{kurtzer2017singularity}. Containerisation also aligns with broader Quick Tips guidance on portable computational environments~\cite{moreau2026containerization}.

Declare containers at the workflow level where possible. For example, Snakemake rules can specify container images for individual steps~\cite{moelder2021snakemake}, and similar patterns exist in other workflow engines. On managed clusters, check whether the centre provides base images configured for the local scheduler, filesystem, MPI stack, GPU drivers, and security policy.

For example, an AlphaFold2--GROMACS workflow may require compatible Python, JAX, CUDA, C++ alignment tools, MPI, and molecular-simulation libraries. Encapsulating these dependencies in an Apptainer image makes the workflow easier to rerun and audit, provided host drivers and accelerator interfaces remain compatible.

\section*{Tip 8: Automate experiment tracking and metadata logging}

AI-HPC workflows often evaluate many model configurations, datasets, random seeds, and software environments. Manual tracking with spreadsheets, flat files, or ad hoc text logs makes it difficult to determine which configuration produced a given result.

Treat model artefacts, metrics, configurations, dataset versions, code commits, random seeds, and execution context as primary workflow outputs rather than incidental by-products. These records should be captured programmatically during training and inference, alongside system-level information such as node type, accelerator use, library versions, and container image identifiers.

Experiment-tracking tools can help link parameters, metrics, artefacts, and execution context across runs. MLflow provides one example of this approach to managing the machine-learning lifecycle~\cite{zaharia2018mlflow}. In digital pathology, for example, a vision-transformer workflow may sweep learning rates, dropout settings, and patch sizes across many whole-slide images. Recording validation curves, commit hashes, dataset versions, and model artefacts for each run makes the selected model auditable and easier to validate externally.

\section*{Tip 9: Design feedback loops explicitly}

AI-driven workflows often contain feedback loops in which intermediate outputs alter later computation. A model may generate predictions that are validated by simulation, with the results then used to update the model, select new inputs, or terminate unproductive branches.

Whereas Tip 1 concerns the placement of AI components within the workflow architecture, this tip concerns the control logic by which model or simulation outputs alter subsequent execution. Feedback should therefore be encoded explicitly rather than managed through manual decisions, disconnected wrapper scripts, or undocumented intervention. The workflow should support iteration, conditional execution, state tracking, and reuse of intermediate results where appropriate.

In directed protein evolution, for example, a generative model may propose sequence variants, rapid screening may prioritise candidates, and selected variants may be routed to higher-fidelity molecular dynamics simulations. The resulting scores or trajectories can then guide the next design--simulation--retraining cycle. Encoding this loop explicitly keeps adaptive execution auditable, reproducible, and easier to orchestrate using the mechanisms discussed in Tip 12.

\section*{Tip 10: Balance single-job performance with workflow throughput}

Single-job benchmarking remains important, but AI-enabled HPC workflows should also be assessed at the workflow level. Relevant metrics include runtime, queue wait time, accelerator utilisation, energy use, successful outputs per unit compute, and reproducibility. A task that runs marginally faster on specialised hardware may still delay the overall workflow if those resources are scarce or heavily queued.

Optimise for end-to-end time to useful scientific result. Use resource requests that allow flexible scheduler placement, batch tasks to keep nodes occupied, and reserve specialised hardware for stages that require it. In iterative workflows, AI components can also reduce unnecessary computation by prioritising candidates likely to improve the model or answer the scientific question, rather than exhaustively executing every possible task.

In a consensus-docking campaign across millions of compounds, extensive optimisation of a single docking kernel may be less valuable than running many moderately sized jobs on readily available CPU partitions, with GPU scoring reserved for selected poses. This improves throughput without dismissing the need to profile and tune genuinely performance-critical kernels.

\section*{Tip 11: Implement appropriate data governance, security, and compliance protocols}

AI-enabled life-science workflows may process identifiable genomic, clinical, or proprietary data. Default permissions on multi-tenant clusters are rarely sufficient when raw inputs, intermediate files, logs, checkpoints, and trained models may all contain sensitive information.

Define governance controls as part of the workflow design. Use site-approved access controls, such as project permissions or POSIX ACLs; encrypt data at rest and during transfer where required; restrict logs and checkpoints; apply retention policies; and avoid copying raw data to uncontrolled scratch locations. Model artefacts should also be treated as sensitive derived data: membership-inference attacks can reveal whether particular records contributed to model training, making governance of trained weights, checkpoints, and logs a practical security requirement~\cite{hu2022membership}.

For example, a clinical workflow analysing rare-disease variants from whole-exome sequencing \texttt{.vcf} files should keep raw genomic data within controlled project storage or site-approved containerised execution environments. Transient arrays may be written to encrypted local scratch or memory-backed storage such as \texttt{tmpfs}, where policy permits, and cleared after completion. Only auditable outputs, provenance records, and appropriately reviewed model artefacts should be retained or shared.

\section*{Tip 12: Use a workflow engine to orchestrate AI--HPC pipelines}

As AI-HPC workflows acquire dependencies, heterogeneous resources, retries, and feedback loops, bespoke shell scripts become brittle. Use a workflow engine to express tasks, inputs, outputs, resource requirements, containers, retries, and provenance explicitly. This connects the feedback logic described in Tip 9 to schedulable execution, rather than relying on manual resubmission or scheduler-log parsing.

Choose the workflow engine according to the execution model. Pipeline-oriented workflows may use systems such as Nextflow, Snakemake, or the Common Workflow Language (CWL)~\cite{ditommaso2017nextflow,moelder2021snakemake,amstutz2016cwl}; more dynamic or task-based workloads may require systems such as Parsl, Pegasus, FireWorks, Dask, or Ray~\cite{babuji2019parsl,deelman2015pegasus,jain2015fireworks,rocklin2015dask,moritz2018ray}. For federated genomics or clinical workflows, standardised task-execution interfaces can support ``bring compute to the data'' designs across HPC, HTC, cloud, and hybrid environments~\cite{kanitz2024ga4gh}. Workflow definitions, environments, metadata, and provenance should also be treated as reusable research objects, consistent with FAIR workflow guidance and recent workflow-community priorities around AI-HPC convergence, heterogeneous resources, and multi-facility execution~\cite{wilkinson2025fairworkflows,ferreiradasilva2024frontiers,ferreiradasilva2024summit}.

In structural biology, for example, a workflow engine can coordinate sequence alignment, deep-learning structure prediction, and molecular refinement across hundreds of proteins. If a GPU prediction task fails, the engine can rerun that task without restarting the whole pipeline, while preserving the dependency graph, outputs, logs, and execution provenance.

\section*{Conclusion}

Many AI-enabled HPC workflows extend relatively linear, deterministic pipelines with adaptive, data-driven stages. While many established HPC practices remain relevant, the integration of AI introduces new challenges in workflow design, data management, and execution.

By applying the principles outlined in these twelve tips, practitioners can design workflows that are more efficient and verifiable, enabling more effective use of modern HPC infrastructure. Together, these approaches support the development of adaptive workflows that can respond dynamically to data, models, and intermediate results. Although the examples in this article are drawn primarily from computational biology and the life sciences, AI-driven HPC workflows in other scientific domains may introduce additional constraints that require domain-specific adaptation of these principles.

%
%
%

\end{document}